\documentclass[12pt]{article}
\usepackage{geometry}
\usepackage{a4}
\usepackage{graphicx}
\usepackage{epsf}
\usepackage{amsmath}
\usepackage{amssymb}
\usepackage{braket}
\usepackage{cite}
\newcommand{\be}{\begin{equation}}
\newcommand{\ee}{\end{equation}}

\newcommand{\Rmnum}[1]{\expandafter\@slowromancap\romannumeral #1@}
\newcommand{\bea}{\begin{eqnarray}}
\newcommand{\eea}{\end{eqnarray}}

\begin{document}
\def\A{{\mathbb{A}}}
\def\B{{\mathbb{B}}}
\def\C{{\mathbb{C}}}
\def\R{{\mathbb{R}}}
\def\s{{\mathbb{S}}}
\def\T{{\mathbb{T}}}
\def\Z{{\mathbb{Z}}}
\def\W{{\mathbb{W}}}
\begin{titlepage}
\title{Information Geometry and the Renormalization Group}
\author{}
\date{
Reevu Maity, Subhash Mahapatra, Tapobrata Sarkar
\thanks{\noindent E-mail:~ reevu, subhmaha, tapo @iitk.ac.in}
\vskip0.4cm
{\sl Department of Physics, \\
Indian Institute of Technology,\\
Kanpur 208016, \\
India}}
\maketitle
\abstract{
\noindent
Information theoretic geometry near critical points in classical and quantum systems is well understood for exactly solvable systems. Here we show that 
renormalization group flow equations can be used to construct the information metric and its associated quantities near criticality, for both classical
and quantum systems, in an universal manner. We study this metric in various cases and establish its scaling properties in several generic examples. 
Scaling relations on the parameter manifold involving scalar quantities are studied, and scaling exponents are identified. The meaning of the scalar curvature 
and the invariant geodesic distance in information geometry is established and substantiated from a renormalization group perspective. 
}
\end{titlepage}
\section{Introduction}
Information geometry provides a unique arena where geometric notions are applied to physical systems, often leading to new and alternative insights into the physics
of classical and quantum phase transitions. A Riemannian metric defined on the parameter space for classical systems or the space of coupling constants for quantum 
systems define a distance on the parameter manifold (PM) \cite{rupp},\cite{pv}. Geometric properties of this distance translate into useful physical quantities 
to understand phase transitions. Although this method is primarily used to study second order continuous transitions, first order phase transitions can also be treated in 
the geometric framework \cite{tapo1},\cite{tapo2}. 

Geometric methods have often been applied to statistical systems that are solvable. Namely, one calculates the metric 
and its associated quantities analytically using an equation of state or a solvable Hamiltonian, and studies their limiting behaviour as one 
approaches criticality. For example, in classical liquid-gas or magnetic systems, one can use the Van der Waals equation of state to analytically compute the 
metric, or use experimental data (based on multi parameter fits to equations of state) to compute the same \cite{tapo1}, \cite{tapo2}. In the context of quantum systems, one normally
alludes to a complex quantum geometric tensor whose real (symmetric) part is the Riemannian metric on the parameter manifold and its imaginary part is the Berry curvature. 
In exactly solvable quantum systems, a knowledge of the ground state leads to the metric \cite{zan1}. 

Scaling analyses of metrics in quantum phase transitions were first
performed in the important work of \cite{VenZan}. Here, the authors provided an integral representation of the quantum geometric tensor in terms of imaginary time 
correlation functions and hence were able to extract information regarding the scaling properties of the metric tensor (see also \cite{Kolo} for related work regarding the
scaling behaviour of the Gaussian curvature in the context of the XY spin chain model). The broad issue that we address in this work is whether there is a generic 
way of understanding the geometry of phase transitions, both classical and quantum, particularly in situations where an exact solution to the model (or an equation of state) 
may not be available. 

To this end, we first note that there are indications that as far as geometry is concerned, the descriptions of classical and quantum phase transitions might
be very similar. Let us briefly elaborate on this by focusing on two dimensional PMs which will mainly be of our interest here. Important in the study of any geometric setup
are scalar invariants in these. These invariants, which are coordinate independent, provide an invariant characterisation of a curved manifold. As is well known, in two dimensions, 
the curvature scalar (or the Ricci scalar) completely characterises the curvature. Associated to this is the scalar expansion parameter (to be elaborated towards
the end of this section) which measures how geodesics (which are analogues of straight lines in curved spaces) converge (or diverge) towards a point in the PM \cite{geod}. 
Further, the line element, identified with an affine parameter that measures infinitesimal distances along geodesics provide a third scalar quantity. 
As pointed out in \cite{tapo3}, Relations between these scalar quantities reveal universal behaviour in classical and quantum phase transitions. 
Namely, the scaling behaviour of the Ricci scalar and the expansion parameter with the  affine parameter near criticality is universal 
for these, in any two dimensional PM, under the assumption that the scalar curvature diverges at criticality as a power law. 
This hint of universality naturally leads one to suspect that there might be a generic way to 
compute metrics on the PM, at least near criticality and we indicate how this can be achieved, by using ideas from scaling symmetries near critical points. 

As far as classical phase transitions are concerned, the usefulness of this method is that we are able to compute the information metric in varied scenario, 
which we believe were hitherto not known. For example, 
as we elaborate upon later, the metric for Ising type models close to, and at, four dimensions computed from our method shows interesting and non-trivial 
behaviour of scalar invariants in information geometry, consistent with the physics near their fixed points and we will show that the scaling behaviour of the
Ricci scalar acquires logarithmic corrections in this example. 

Although most of this work deals with classical phase transitions, we apply our method to one example in the context of zero temperature quantum phase
transitions and compute the metric using scaling arguments, and show that we get consistent results. Our method here should be contrasted with the one 
developed in \cite{VenZan}. As we have mentioned, the latter used the imaginary time correlation function to derive the scaling relations of the metric in the 
context of quantum phase transitions. Here, we will directly appeal to the renormalization group (RG) equations to construct the metric and show that this 
gives sensible results. In a sense this complements the approach of \cite{VenZan}, while retaining universal applicability. 

It is useful to point out here the conventional definitions of the information metric in classical and quantum systems. In the former case, Ruppeiner's
definition of the metric \cite{rupp} reduces to derivatives of a thermodynamic potential. For example, if we consider the entropy density $s$, then the line element and the 
metric on the PM are defined via 
\begin{equation}
d \lambda^2 = g_{ab}dx^adx^b, ~~~g_{ab} = -\left(\frac{\partial^2 s}{\partial x^a \partial x^b}\right),
\label{cmetric}
\end{equation}
where $x^{a}, a=1,2$, denotes the internal energy and the particle number (both per unit volume) and the Boltzmann's constant is set to unity. 
These are the co-ordinates on the parameter manifold in the ``entropy representation.''  The metric can also be computed in various
other representations and a full list of such metrics is available in Table II of \cite{rupp}. 
We mention here that an alternative definition of the metric had been advocated earlier in the seminal works of Weinhold \cite{Weinhold}. 
That the two definitions are related by a conformal transformation is well known. 

On the other hand, the information metric in quantum systems is defined by considering two infinitesimally separated quantum states 
(in the parameter space) and computing 
\begin{equation}
|\psi\left({\vec x} +d{\vec x}\right) - \psi\left({\vec x}\right)|^2 = \langle \partial_{a} \psi|\partial_{b} \psi\rangle dx^{a}dx^{b} = \alpha_{a b}dx^{a}dx^{b}
\label{pv1}
\end{equation}
where $x^{a}$ (collectively denoted as ${\vec x}$ in the l.h.s of eq.(\ref{pv1})) denotes the parameters on which the wave function $\psi$ depends on, and
$\partial_{a}$ is a derivative with respect to $x^{a}$. 
The approach of \cite{pv} is to construct, from $\alpha_{a b}$ (which are not gauge invariant), a gauge-invariant metric tensor given by
\begin{equation}
g_{a b} = \alpha_{ab} - \beta_{a}\beta_{b};~~~~\beta_{a} = -i\langle\psi\left({\vec x}\right)|\partial_{a}\psi\left({\vec x}\right)\rangle
\label{qmetric}
\end{equation}

Here, $g_{a b}$ is the metric induced on the PM from the natural structure of the Hilbert space of quantum states. Eqs.(\ref{cmetric}) and (\ref{qmetric}) are
the standard definitions of the classical and quantum mechanical Riemannian metrics on the parameter manifold. Our approach here would be to compute these metrics 
in the critical regime, without using Eqs.(\ref{cmetric}) and (\ref{qmetric}) directly. 

For this purpose, we use an existing notion in the literature, namely the geometric equivalent of scale invariance near a fixed point. To the best of our
knowledge, such a proposal first appeared in \cite{Diosi}. In this work, the scaling relations for classical liquid-gas phase transitions are recovered from a geometric 
perspective. Related work has appeared in the literature in the context of quantum field theory \cite{Lassig}, \cite{Dolan1}, \cite{Dolan2}, \cite{Sayan1}, \cite{Heckman1},
\cite{Heckman2} and statistical mechanics \cite{BrodyRitz}. The main idea in these works is that the renormalization group flow equations determine the so called 
homothetic vector fields, which are mathematically related to scale invariance. 

Specifically, if $K^a$ is a homothetic vector field on a manifold with metric $g_{ab}$ 
(we will mostly consider two dimensional manifolds so that $a,b=1,2$), it satisfies the condition ${\mathcal L}_K g_{ab} = Dg_{ab}$, 
where ${\mathcal L}_K g_{ab}$ is the Lie derivative (see, e.g section 1.4 of \cite{poisson})
of the metric along a curve whose tangent is $K^a$. Also $D$ is a constant that we will identify with the spatial dimension of the system (note that this 
is different from the dimensionality of the parameter manifold, which will mostly be two in this paper). This equation reduces by standard manipulations to
the condition $K_{a;b} + K_{b;a} = Dg_{ab}$ where a semicolon denotes a covariant derivative (defined in the next section) on the parameter manifold (again, 
see section 1.4 of \cite{poisson}). 
This last equation can be alternatively written in a simpler form as  
\begin{equation}
g_{ac}K^{c}_{,b}+g_{bc}K^{c}_{,a}+g_{ab,c}K^{c}=Dg_{ab},
\label{homothety}
\end{equation}
where the comma indicates an ordinary derivative with respect to the coordinate label that follows it, and repeated indices imply a summation (which will always
be the case in this paper). To fix ideas, let us consider a textbook example, 
the two dimensional Euclidean space. This is flat space, with coordinates $(x,y)$, and metric $g_{ab} = {\rm diag}(1,1)$. Considering the vector field $K^a = (x,y)$, 
which represent the tangent at any point on the flat manifold, it is seen that 
$K_{a;b} + K_{b;a}= {\rm diag}(2,2)$, thus confirming Eq.(\ref{homothety}). This is expected, since two dimensional Euclidean space is flat, and looks the
same at any length scale. Curved manifolds, which will be of our interest here are more challenging to deal with. Indeed, properties of homothetic
vectors (when they exist) are of great importance in General Relativity and Cosmology. Here, we will apply this to information geometry of phase transitions.  

In this paper, we will use the notion of scale invariance of the parameter manifold near a critical point. 
Our starting point is a metric on the parameter manifold of a system, i.e the space of the coupling constants in the theory. This metric is assumed to 
be a priori unknown. Near criticality, following \cite{Diosi}, \cite{Lassig}, we demand that the beta functions of the theory are the components of 
a tangent vector field which is homothetic. From Eq.(\ref{homothety}), we then get a set of coupled partical differential equations for the components of
the metric. These equations, if solvable, will lead to solutions of the metric on the PM, without a detailed knowledge of the full solution of the system.

For classical systems where, in the conventional approach of Ruppeiner \cite{rupp}, the metric components are defined via derivatives of the free energy 
(or entropy) and are related to response functions, this has been demonstrated in \cite{Diosi}. Here, up to linear order in RG, it was found that 
Eq.(\ref{homothety}) implied that the metric components are generalized homogeneous functions. Euler's theorem was then invoked to read off the
scaling behaviour of the metric components. In this paper, we will consider situations where this may not be possible, and solve for the metric components
directly from Eq.(\ref{homothety}). Using scale 
invariance of the parameter manifold, our method should be viewed as a tool for obtaining the geometry of any system, 
sufficiently near to criticality. This non-trivially generalizes the analysis of \cite{Diosi} by providing an universal approach to computing 
such metrics, and as we show in sequel, we obtain novel properties of the information metric for a wide variety of systems, consistent with the physics near
the fixed points of these. It should be kept in mind that if, in general, the coordinates on the coupling constant space are $(x,y)$,
while linearizing about a non-trivial fixed point $(x^*,y^*)$, it is more natural to use as coordinates $(\delta x, \delta y) = (x-x^*, y-y^*)$. This should
be understood by the context. 

Viewed in this perspective, this method bypasses the standard requirement of the knowledge of the equation of state (for classical systems) or the
many body ground state (for quantum systems). Of course this is assuming that the system of equations generated from Eq.(\ref{homothety}) are solvable,
but we will show in sequel that this is true in a variety of examples. An objection could be that even if Eq.(\ref{homothety}) does yield a solution up to
some level in RG, this might not be the case when higher order terms are introduced. This is certainly a drawback, which we will address towards the 
end of this paper. For most of this paper, we will concentrate on cases where Eq.(\ref{homothety}) admits an analytic solution. We will see in sequel that
we are able to capture a large class of models within this simple minded approach. 

As mentioned earlier, important ingredients in any geometric setup are scalars, which are invariant under coordinate 
transformations and indicate global properties of a curved manifold. In contrast, tensor components such as metric components will change under a coordinate
transformation, and their scaling relations will not in general be coordinate invariant. Ruppeiner conjectured \cite{rupp}
that near criticality, $R \sim \xi^D$. The arguments of Ruppeiner are based on the notion of 
relative flatness in a curved space (see Eq.(4.76) of \cite{rupp} and the arguments preceeding this equation). From an RG perspective, we will prove that this result
is exact up to linear order, but that there are important subtleties when one includes a class of higher order terms. Further, the 
infinitesimal geodesic distance $d\lambda$ along a curve, defined (see Eq.(\ref{cmetric})) as $d\lambda = \sqrt{g_{ab}dx^adx^b}$ is an interesting 
quantity and is known to be related to the concept of a 
statistical distance. We study this object, and show that the geodesic distance is related to the length scale of the problem. 

It is well known that that geodesics converge (or diverge) at singularities of a given manifold \cite{geofocus}. From a celebrated equation 
due to Raychaudhuri  (see e.g \cite{sayan}), the convergence of geodesics can be quantified in this case by a scalar expansion parameter (called $\Theta$ in sequel). 
In two dimensional Euclidean cases which are of our interest here, the expansion parameter is also an universal indicator of phase transitions, 
as is the Ricci scalar.  There are algebraic relations between the three scalar quantities mentioned above, defined on a two dimensional manifold. 
These give rise to the so called geometric exponents \cite{tapo3}. While in that paper, these exponents were calculated in solvable systems, we show here 
that they emerge from the perspective of RG, even beyond linear order. 

This paper is organized as follows. In section 2, we elaborate on linearized RG flow equations, generalizing the work of \cite{Diosi}, and providing a number
of new results. In section 3, we study non-linear RG flow equations and their geometric significance, including logarithmic corrections. Section 4 ends this paper
with our conclusions and directions for future research. 

Before we embark on our analysis, a word about the notations and conventions used in the paper is in order. We consider a variety of examples, and 
using different symbols for the variables will unnecessarily clutter the notation. We will proceed with the understanding that the notations used in a
particular subsection of this paper find usage only in that subsection and are not to be related to the other subsections of the paper. Also, the examples used
in this paper are standard and can be found in textbooks \cite{Goldenfeld}, \cite{Cardy}. We will refrain from a detailed discussion of the models
themselves, which will make the paper unnecessarily lengthy, and instead refer the reader to these excellent texts for more details. 
Another important issue should be kept in mind. In order to have a valid notion of geometry, the line element 
$d\lambda^2 = g_{ab}dx^adx^b$ should be
positive definite. This means that along with the diagonal element, the determinant of the metric tensor should be positive (see, e.g the discussion around
Eq.(3.20) of \cite{rupp}). In all the examples considered in this paper, we have checked that this condition is satisfied. We will not mention this in sequel. 

\section{Linearized RG and information geometry}

In this section, we demonstrate the construction of the information theoretic metric near criticality for linearized RG flows. These may arise in any 
statistical system, when one linearizes the RG equations near a critical point. Here, we will be concerned with two parameter examples, 
and comments on the generalization to higher dimensional manifolds will be given towards the end of this section. 

\subsection{Case I}

In this subsection, we first recast the results of \cite{Diosi} in a form that will be useful for us to make some general statements regarding two parameter
information theoretic models in linear RG, and then go on to study geodesics for these models. 
We start with a theory with two coupling constants $x$ and $y$, and assume that near a generic fixed point $(x^*,y^*)$, 
the linearized RG equations can be written in terms of the eigenvalues $a$ and $b$ (these should not be confused with the coordinate labels) as
\begin{equation}
{\dot x} = ax, ~~~{\dot y} = by,
\label{linear}
\end{equation}
where the overdot indicates a derivative with respect to a logarithmic length scale, $l = {\rm ln}(L)$. As pointed out in the introduction, appropriate coordinates
on the PM in this case are $(\delta x, \delta y) = (x-x^*, y-y^*)$, which we will still call $(x,y)$ by a slight abuse of notation. Let us denote the metric components 
on the PM by $g_{xx},~g_{yy}$ and $g_{xy}$ \cite{diag}.

As mentioned in the introduction, for this example the homothetic vector field has components $K^a = (ax,by)$ \cite{Hom1}. 
We take this as an input and insert it in Eq.(\ref{homothety}) (or, alternatively use $K_{a;b} + K_{b;a} = Dg_{ab}$ after lowering the indices of $K^a$). 
Writing out the components of Eq.(\ref{homothety}) then leads to differential equations for the metric tensor. For the chosen off-diagonal form of the metric, 
Eq.(\ref{homothety}) gives rise to the following three equations (as before, a comma denotes an ordinary derivative with respect to the variable that follows it) :
\begin{eqnarray}
&~&axg_{xx,x} + byg_{xx,y} + (2a - D)g_{xx} = 0~, \nonumber\\
&~&axg_{yy,x} + byg_{yy,y} + (2b - D)g_{yy} = 0~, \nonumber\\
&~&axg_{xy,x} + byg_{xy,y} + (a + b - D)g_{xy} = 0~.
\label{linone}
\end{eqnarray}
This reveals that the metric components are generalized homogeneous functions near criticality \cite{Diosi} and immediately
reproduces the well known static scaling relations. This is true for any linearized set of RG equations. Say the variable $x$ drives the phase transition. 
Then we can write the general solution for Eq.(\ref{linone}) as 
\begin{equation}
g_{xx} = x^{\frac{D}{a}-2}{\mathcal G}_1\left(y x^{-\frac{b}{a}}\right),~~
g_{yy} = x^{\frac{D-2 b}{a}} {\mathcal G}_2\left(y x^{-\frac{b}{a}}\right),~~
g_{xy} = x^{-\frac{a+b-D}{a}} {\mathcal G}_3\left(y x^{-\frac{b}{a}}\right).
\label{metriclin}
\end{equation}
Here, ${\mathcal G}_i,~i=1,2,3$ are functions of a single variable $y x^{-\frac{b}{a}}$, and reminiscent of Widom scaling of the free energy
near criticality \cite{Goldenfeld}, \cite{Cardy}. However, the functions ${\mathcal G}_i$ are not the same as scaling functions that apper in the free
energy, since the metric is more naturally interpreted as the second derivative of the free energy in the conventional picture of information geometry. 
It will be assumed that the functions ${\mathcal G}_i$ are analytic and equal a constant value near criticality, $y=0$. We will advocate two arguments 
to justify this, and go on to check these with the known example of the 1-D Ising model. 

First, note that in the classical notion of information geometry, the metric components  are related to the classical response functions. For example, 
in a magnetic system, if $x$ is identified with the reduced temperature $t = (T - T_c)/T_c$, with $T$ being the temperature and $T_c$ its critical value, and $y$ is 
identified with the reduced magnetic field $H/T_c$, then the metric components $g_{xx}$, $g_{yy}$ and $g_{xy}$ are related to the specific heat, the magnetic 
susceptibility and the derivative of the magnetization, respectively. Eq.(\ref{metriclin}) then indicates that these have the correct critical exponents if the
functions ${\mathcal G}_i$ are analytic at $y = 0$, and equal a constant of order unity. 

This can also be seen by noting that Eq.(\ref{linone}) translates into the fact that the metric components are generalized homogeneous functions up to linear order 
so that standard scaling arguments can be applied (see Eq.(5.5) of \cite{Diosi}). This is indicative of the fact that the functions ${\mathcal G}_i$ 
can be taken to be analytic, and constants of order unity close to criticality. There is a small subtlety here. If we take all functions ${\mathcal G}_i$ to 
equal the same constant near criticality, the metric of Eq.(\ref{metriclin}) become singular. Hence, this should be avoided, and ${\mathcal G}_i, i=1,2,3$ 
have to be taken to equal different constant numbers of order one. These multiplicative constants can at most affect our results for the metric, the scalar curvature 
and the expansion parameter by some overall constants, and will not affect our scaling analysis. Keeping these explicitly in the computations will clutter
the notation, and without loss of generality, we will take two of these to equal unity, and set the other one to $2$. This is just
a particular choice and any other choice would only affect the results by an overall numerical constant. With this choice, we are also able to consistently 
satisfy the positivity constraint on the line element, as can be checked. 

Secondly, assuming that the scaling function is analytic, and equals a constant of order unity 
near criticality, we obtain the scalar curvature in terms of the driving parameter $x$ up to an overall constant as 
\begin{equation}
R = \frac{2 b}{a^2} (D-2 b) x^{-\frac{D}{a}}
\label{Rone}
\end{equation}
This shows that the scalar curvature blows up if $x$ is relevant, and goes to zero of $x$ is irrelevant. If we assume that 
$R \sim \xi^D$ where $\xi$ is the correlation length, then Eq.(\ref{Rone}) implies that $\xi \sim x^{-\frac{1}{a}}$, i.e correctly 
reproduces the correlation length exponent for classical phase transitions where $x$ is identified with the reduced temperature. 
For quantum phase transitions, we note that if $x$ is a relevant variable, then a perturbation in this direction produces a gap in the spectrum 
that in turn indicates that the correlation length scales as $x^{-\frac{1}{a}}$. This is again consistent with $R \sim \xi^D$ (and also justifies the
assumption of $R \sim \xi^D$). We thus see that assuming that the functions ${\mathcal G}_i$ are constants of
order unity near criticality produces a consistent geometric picture up to linear order in RG. This will be assumed in what follows. We will
not explicitly indicate these functions in sequel. 

It is instructive to validate our analysis thus far by comparing it to a known example. We choose the standard example of the classical 1-D Ising model 
in a magnetic field, with the Hamiltonian given by 
\begin{equation}
H = -J\sum_{j=1}^N S_jS_{j+1} -h\sum_{j=1}^N S_j
\end{equation}
Information geometry for this model was worked out in \cite{JanMrug} in the limit of large $N$, and we quote their result for the metric. First we define the variables
$x = J/T$ and $y = h/T$ (where we set the Boltzmann's constant to unity). Further, writing $t = e^{-4x}$ and near the critical point substituting $t = \epsilon$ and $h = \delta$,
the metric components (Eq.(4.17) of \cite{JanMrug}) read, after some algebra, 
\begin{equation}
g_{xx} = \frac{\epsilon^{-\frac{3}{2}}}{4}\frac{1+2(\delta\epsilon^{-\frac{1}{2}})^2}{\left(1+(\delta\epsilon^{-\frac{1}{2}})^2\right)^{\frac{3}{2}}},~~
g_{xy} = \frac{\epsilon^{-1}}{2}\frac{\delta}{\epsilon^{\frac{1}{2}}}\frac{1}{\left(1+(\delta\epsilon^{-\frac{1}{2}})^2\right)^{\frac{3}{2}}},~~
g_{yy} = \epsilon^{-\frac{1}{2}}\frac{1}{\left(1+(\delta\epsilon^{-\frac{1}{2}})^2\right)^{\frac{3}{2}}}
\end{equation}
Thus the metric is similar to the one in Eq.(\ref{metriclin}) (with $x \equiv \epsilon$ and $y \equiv \delta$) 
upon identification $a=2$, $b=1$, with $D=1$. The functions defined in that equation read
\begin{eqnarray}
{\mathcal G}_1\left(\delta \epsilon^{-\frac{1}{2}}\right) &=&  \frac{1}{4}\frac{1+2(\delta\epsilon^{-\frac{1}{2}})^2}{\left(1+(\delta\epsilon^{-\frac{1}{2}})^2\right)^{\frac{3}{2}}},~~
{\mathcal G}_2\left(\delta \epsilon^{-\frac{1}{2}}\right) = \frac{\delta\epsilon^{-\frac{1}{2}}}{2\left(1+(\delta\epsilon^{-\frac{1}{2}})^2\right)^{\frac{3}{2}}}\nonumber\\
{\mathcal G}_3\left(\delta \epsilon^{-\frac{1}{2}}\right) &=& \frac{1}{\left(1+(\delta\epsilon^{-\frac{1}{2}})^2\right)^{\frac{3}{2}}}
\end{eqnarray}
which are analytic near criticality, as expected, if we assume $\epsilon$ and $\delta$ to be of the same order. Note that these are different functions which
evaluate to different numerical values near criticality, as alluded to before. Also, from our discussion it follows that 
the RG flow equations here are governed by ${\dot t}=2t$, ${\dot h} = h$, in agreement with Eq.(32) and (33) of \cite{BrodyRitz}. The analysis of the
scalar curvature and geodesics for the one dimensional Ising model has been done in \cite{tapo3}, to which we refer the reader for more details.  

The metric of Eq.(\ref{metriclin}) is to be used when $x$ is the driving parameter in the phase transition. 
An equivalent form of writing the solutions of Eq.(\ref{linone}) is
\begin{equation}
g_{xx} = y^{\frac{D-2a}{b}}{\mathcal F}_1\left(x y^{-\frac{a}{b}}\right),~~
g_{yy} = y^{\frac{D-2 b}{b}} {\mathcal F}_2\left(x y^{-\frac{a}{b}}\right),~~
g_{xy} = y^{-\frac{a+b-D}{b}} {\mathcal F}_3\left(x y^{-\frac{a}{b}}\right).
\label{metriclintwo}
\end{equation}
Here ${\mathcal F}_i,~i=1,2,3$ are arbitrary functions of the variable $x y^{-\frac{a}{b}}$ and are assumed to approach a constant value
at $x = 0$. The metic of Eq.(\ref{metriclintwo}) should be used when $y$ drives the phase transition. This metric has a scalar curvature given by 
\begin{equation}
R = \frac{a}{b^2} (D-2 a) y^{-\frac{D}{b}}
\label{Rtwo}
\end{equation}
Comparing Eqs.(\ref{Rone}) and (\ref{Rtwo}), we see that the divergence of the scalar curvature is controlled by the coefficient of the driving parameter
in the RG, which is expected. 
As before, the functional forms of ${\mathcal F}_i\left(x y^{-\frac{a}{b}}\right)$ 
may be different for $i=1,2,3$. However the only assumption here is that all these are of order unity and that the leading behaviour of the metric near criticality
is controlled by the exponents of $y$.  

Eq.(\ref{Rone}) is applicable to any model of linearized RG and shows that if the parameter $x$ is relevant, the divergence of the scalar curvature is controlled
by the relevant eigenvalue. If we identify the scalar curvature with the correlation volume (up to a possible arbitrary constant), it is seen that the correct 
correlation length exponent is recovered.  

An interesting quantity that we will now focus our attention on is a set of geodesics (called a geodesic congruence) on the PM. Geodesics are analogues of straight lines
in curved spaces, and these are paths that minimize distances between points on a curved manifold. For curved information theoretic manifolds that we 
describe here, geodesics provide a further characterization of classical and quantum phase transitions, as shown in \cite{tapo4}. Namely, one considers a geodesic
congruence on the PM, and it can be shown that near a critical point, the congruence converges (or diverges). 

Let us make this statement more precise. If our PM is defined by
the coordinates (i.e coupling constants) $x^{a}$, then geodesic paths on the manifold satisfy the equation 
$(x^{a})^{''} + \Gamma^{a}_{bc}(x^{b})^{'}(x^{c})^{'}
= 0$. Here, $\Gamma^{a}_{bc} = \frac{1}{2}g^{ad}\left(g_{db,c}+ g_{dc,b}-g_{bc,d}\right)$ are the Christoffel 
connections defined from the metric, and the prime denotes a derivative with respect to an affine parameter $\lambda$ along the geodesic, which is conventionally
taken to be the square root of the line element, i.e $d\lambda^2 = g_{ab}dx^{a}dx^{b}$. For such an affinely parametrized 
geodesic, the geodesic equations can be obtained from a variational principle from the Lagrangian ${\mathcal L} = \frac{1}{2}\left(g_{ab}(x^{a})^{'} (x^{b})^{'}\right)$. 
This will be useful for us later. 

If we denote as the normalized tangent vectors $u^{a} = (x^{a})^{'}$, curvature effects on geodesics (near criticality) 
are measured by the tensor $B^{a}_{\phantom{1}b}=\nabla_{b}u^{a}$. 
Recall that the covariant derivative on a generic vector $V^{a}$ is defined by the action $\nabla_{a}V^{b} = \partial_{a}V^{b} + \Gamma^{b}_{ac}V^{c}$.
Then it can be shown that $\Theta = B^{a}_{\phantom{1}a}$, called the expansion scalar, gives an effective measure of the convergence or divergence
of a geodesic congruence. At critical points on the PM, i.e at
phase transitions, $\Theta$ diverges.  $\Theta$ being a scalar quantity, this is a coordinate independent characterization of phase transitions. 
To compute $\Theta$, we require a solution for the vectors $u^a$. In general this might be difficult to obtain, but when the metric is independent
of one of the coordinates (as will always be the case here), such solutions can be found analytically if $u^a$ is normalized, i.e $u^au_a=1$. These
analytic solutions constitute our geodesic congruence. We refer the reader to \cite{tapo3} for more details. 

To illustrate the procedure, we first recall the two dimensional metric of Eq.(\ref{metriclin}), where we will include a multiplicative factor of 
$k_1$ in the $g_{yy}$ component. Now denote the tangent vectors to geodesic trajectories
by the vector $u^{a} = \left(x'(\lambda),y'(\lambda)\right)$, where a prime denotes a derivative with respect to an affine parameter $\lambda$. Normalization 
of $u^{a}$ imposes the condition 
\begin{equation}
x(\lambda )^{\frac{-2 a-2 b+D}{a}} \left[x'(\lambda )^2 x(\lambda )^{\frac{2b}{a}}+2 x'(\lambda ) y'(\lambda ) x(\lambda )^{\frac{a+b}{a}}+
k_1x(\lambda )^2 y'(\lambda )^2\right]=1
\label{norm}
\end{equation}
The left hand side of Eq.(\ref{norm}) is in fact proportional to the Lagrangian alluded to before. Noting that this is independent of the coordinate 
$y$ (as is the metric), the Euler-Lagrange equation for $y(\lambda)$ imposes a further constraint $\partial {\mathcal L}/\partial y' = k_2$, where $k_2$ 
is an arbitrary constant. Then, if this constraint is solved in conjunction with the normalization condition, we obtain
\begin{eqnarray}
x'(\lambda) &=&\sqrt{x(\lambda )^{2-\frac{2 D}{a}} \left(2 x(\lambda )^{\frac{D}{a}}-k_2^2x(\lambda )^{\frac{2 B}{a}}\right)},~~\nonumber\\
y'(\lambda) &=& \frac{1}{2} \left(k_2 x(\lambda )^{\frac{2 B-D}{a}}-x(\lambda)^{\frac{B}{a}-1} \sqrt{x(\lambda )^{2-\frac{2 D}{a}} 
\left(2 x(\lambda)^{\frac{D}{a}}-k_2^2 x(\lambda )^{\frac{2 B}{a}}\right)}\right)
\label{primes}
\end{eqnarray}
The first of these equations can be solved to obtain an expression for the geodesic distance in terms of the Gauss Hypergeometric function,
\begin{equation}
\lambda = \frac{\sqrt{2}}{D} a x^{\frac{D}{2 a}} \, _2F_1\left(\frac{1}{2},\frac{D}{4 b-2D};\frac{D}{4 b-2 D}+1;\frac{1}{2} k_2^2 x^{\frac{2 b-D}{a}}\right)-k_3~,
\label{lam1}
\end{equation}
where $k_3$ is another arbitrary constant. In order to obtain a real value of $\lambda$ which is physically reasonable, we require $b > D/2$, and a further constraint 
on the constant $k_2$. For small values of $x$, this can be seen to restrict the Hypergeometric function to values close to unity.  This last fact indicates that 
it is reasonable to set $k_2 = 0$, without loss of generality. Indeed, Eq.(\ref{primes}) simplifies in this limit and we obtain as a solution,
$\lambda \sim x^{\frac{D}{2a}} - k_3$. Now note that we are interested in geodesics that reach very close to the critical point.  It is natural to measure 
$\lambda$ from the critical point, so that we will require $\lambda \to 0$ as $x \to 0$. This indicates that the constant $k_3$ can be set to zero as well \cite{aneg}. 

From Eq.(\ref{primes}), we now obtain the following solutions for $x$ and $y$ as a function of $\lambda$ :
\begin{equation}
x(\lambda) = 2^{-\frac{a}{D}} \left(\frac{D \lambda }{a}\right)^{\frac{2a}{D}},~~
y(\lambda) = -\frac{2^{-\frac{b+D}{D}}a}{b}\left(\frac{D \lambda}{a}\right)^{\frac{2 b}{D}},
\label{solxy}
\end{equation}
where we have imposed the condition $x(\lambda = 0) = 0$, i.e the affine parameter is measured from criticality. 
These equations can be now inverted to obtain an analytic expression for the
affine parameter, namely, $\lambda \sim x^{\frac{D}{2a}}$, apart from constant factors. Also, using Eq.(\ref{primes}) and the metric of Eq.(\ref{metriclin}), we 
obtain by some simple manipulations, 
\begin{equation}
\Theta = \frac{(D-2 b) x^{-\frac{D}{2 a}}}{\sqrt{2} a}.
\label{thetalin}
\end{equation}
Using the solution for the affine parameter, we obtain as $x\to 0$ from Eqs.(\ref{Rtwo}) and (\ref{thetalin}),
\begin{equation}
R \sim \lambda^{-2},~~\Theta \sim \lambda^{-1}
\label{geoexponents}
\end{equation}
The same conclusion can be reached for the metric of Eq.(\ref{metriclintwo}), as can be easily checked. Eq.(\ref{geoexponents}) can be thus 
understood as an universal indicator of phase transitions for any system with linearized RG flow. The exponents appearing in this equation were 
dubbed as geometric critical exponents in \cite{tapo3}. In that paper, the analysis was conducted by exploiting the behavior of the information
metric close to criticality for exactly solvable systems. Here we have given proof that the relations hold for any arbitrary two parameter system,
at least up to linear order in RG. 

We record a couple of observations before we move on. From Eq.(\ref{Rtwo}) and (\ref{thetalin}), note that the scalar curvature and the expansion
parameter diverge if the operator $x$ (or $y$) are relevant. If these are irrelevant, i.e $a$ or $b$ is negative so that $x$ or $y$ are stable directions, 
then these go to zero in the limit that the coupling constants go to zero. In that case, $\lambda$ calculated from Eq.(\ref{lam1}) (after setting $k_2$ and 
$k_3$ to zero) approach infinity. This is a typical feature of information geometry that we will come across later also \cite{finetune}.  Note that 
the relations of Eq.(\ref{geoexponents}) remain valid irrespective of whether the operators are relevant or irrelevant. 

Also note that the geodesic distance $\lambda$ can be related to the length scale of the problem as follows. Using the fact that under an RG
transformation, $\xi = \xi_0e^{l}$, we find that $R \sim \xi^D $ translates into $l = -(2/D){\rm log}\lambda$. This is a generic feature for all linearized cases. 

\subsection{Case II}

Our next example is that of an RG flow of the form 
\begin{equation}
{\dot x} = a_1x + a_2y, ~~~{\dot y} = b_1y.
\label{mixed1}
\end{equation}
This form of the RG equations occur in perturbation theory, for a linearized one-loop approximation in Landau-Ginzburg models. 
In this case, one can obtain a metric using the original variables, as we illustrate in a moment. The important point is that geometric methods can be
applied to set of redefined coordinates, consistent with the eigendirections of the RG flow equations. For example, in this case, if we define 
a new variable $z = x + a_2y/(a_1 - b_1)$, the RG flow equations reduce to ${\dot z} = a_1z$, ${\dot y} = b_1y$. 
The results of our previous analysis can now be readily applied in this new set of coordinates. In particular, for $y=0$, we obtain the components of the 
information metric as
\begin{equation}
g_{zz} = z^{\frac{D}{a_1}-2},~~ g_{yy} = z^{\frac{D-2 b_1}{a_1}},~~g_{yz} = z^{-\frac{a_1+b_1-D}{a_1}} 
\label{LGone}
\end{equation}
and an entirely similar analysis holds for $z=0$. In both cases it can be seen that our previous results $R \sim \lambda^{-2}$ and $\Theta \sim \lambda^{-1}$ hold. 
As before, the scalar curvature can be computed and the expressions are similar to the ones in Eq.(\ref{Rone}) and (\ref{Rtwo}), and with the identification
$R \sim \xi^D$ shows that while $a_1$ is the critical exponent in the direction $y=0$, $b_1$ is the one in the direction $z=0$. We have thus 
constructed the information metric for the Landau-Ginzburg model at one loop, Eq.(\ref{LGone}), solely by using the RG flow equations. 

We should mention here that using the original set of equations (Eq.(\ref{mixed1})) also, it is possible to compute the metric tensor. However,
this has a complicated structure and does not reveal any meaningful physics. Our scalar relations are however expected to hold here as well. 
It should thus be kept in mind that to interpret the various quantities associated with information geometry, one needs to 
correctly choose coordinates. Once this is done, analysis of the metric becomes meaningful, and the correlation length exponent comes out
correctly with the identification $R \sim \xi^D$, with $D$ being the spatial dimension of the system. To summarize, for any set of linearized 
RG flow equations, the information metric can be written down simply from the scaling dimension of the operators. An appropriate choice of coordinates
then predicts the correct exponents of the system. 

\subsection{Case III}

Before we close this section, we will comment on a situation in which a system has a critical line, for example a gapless line in 
the parameter space for quantum phase transitions. This is exemplified by the one-dimensional anisotropic Heisenberg 
spin $1/2$ chain. This model was considered in \cite{DuttaSen}, where the gapless line was interpreted as a spin flip transition. 
The model Hamiltonian is given by 
\begin{equation}
H = \sum_n \left[(1+\gamma)S^x_nS^x_{n+1} + (1-\gamma)S^y_nS^y_{n+1} +\Delta S^z_nS^z_{n+1} -hS^x_n\right]
\label{Hamxyz}
\end{equation}
where $S^i,~i=1,2,3$ are spin operators, $\gamma$ is an anisotropy parameter, $\Delta$ is the coupling in the $z$-direction and 
$h$ is a magnetic field along the $x$ direction. As shown in \cite{DuttaSen}, bosonization techniques yield the following perturbative RG equations
in terms of $h$, $\gamma$ and $b$ (the latter being the coefficient of an operator which arises in an operator product expansion) are 
\begin{equation}
{\dot h} = a_1h -a_2\gamma h - a_3 bh,~~{\dot \gamma} = b_1 \gamma - b_2 h^2,~~
{\dot b} = c_1b + c_2 h^2~,
\label{DuttaSenRG}
\end{equation}
where the coefficients are determined by the scaling dimension of the corresponding operators and read 
\begin{eqnarray}
a_1 &=& 2-K - 1/(4K),~~a_2=2-1/K,~~a_3=2- 4K\nonumber\\
b_1 &=& (2-1/K),~~ b_2 = (2K - 1/(2K)) = c_2,~~c_1 = (2-4K)
\end{eqnarray}
Here, $K$ is related to $\Delta$ (Eq.(3) of \cite{DuttaSen}) 
and takes values $1/2\leq K < \infty$ (for details, see \cite{DuttaSen}). Now suppose we are at a fixed point $h=h^*$, and look at 
information geometry in the $\gamma - b$ plane, where the RG equations are linear. We will not go into the details here, but simply state the result that the 
information metric at a fixed point of $h$ can be obtained to be 
\begin{equation}
g_{\gamma \gamma} =z^{-2 -\frac{K}{1-2K}},~~g_{\gamma b} =z^{\frac{1 - K(4K-3)}{1-2K}},~~g_{bb} = z^{\frac{K(3-8K)}{1-2K}}
\label{ds1}
\end{equation}
where $z =  \left(2\gamma(1-\frac{1}{2K}) - 2h^{*2}(K-\frac{1}{4K})\right)$, and a coordinate defined by 
$y=b(2-4K) + h^{*2}(2K-1/(2K))$ is set to zero. The scalar curvature and the expansion parameter diverge as 
\begin{equation}
R \sim z^{-\frac{2K}{2K-1}},~~\Theta \sim z^{-\frac{K}{2K-1}}
\end{equation}
and it can be checked that $R \sim \lambda^{-2}$ and $\Theta \sim \lambda^{-1}$, with $\lambda$ being the geodesic length, as expected. 
For different values of $K$ within its specified range, the scalar curvature and the expansion parameter diverges at 
$\gamma^* = (2K+1)h^{*2}/2$. An entirely similar analysis holds for $z=0$ when the information metric is determined by $y$. This results
in $b^* = \gamma^*/(2K)$. These values of $\gamma^*$ and $b^*$ determine a fixed line in the $\gamma - b$ plane and also determine the 
value of $h^{*}$ which is entirely consistent with \cite{DuttaSen}, proving the validity of Eq.({\ref{ds1}). 

As an aside, we point out that a similar RG flow equation as in Eq.(\ref{DuttaSenRG}) was obtained in \cite{Nersesyan} for a model of two
weakly coupled Luttinger chains, where the second and third terms of the first of Eq.(\ref{DuttaSenRG}) were absent. In this case, a simple 
transformation of variables $z_1 = \gamma - b_2 h^2/(2a_1 - b_1)$ and $z_2 =  b - c_2h^2/(2a_1 - c_1)$ renders the RG equations linear,
i.e ${\dot h} = a_1h$, ${\dot z_1}= b_1z_1$, ${\dot z_2} = c_1 z_2$. These equations can be used to define the homothetic vector field in three dimensions,
where an entirely similar analysis as our two dimensional examples so far can be done. We will however not present the details here : higher 
dimensional examples will be treated elsewhere. 

Finally we point out that setting $\Delta = 0$ in Eq.(\ref{Hamxyz}) reduces it to the transverse XY model considered in the context of information 
geometry in \cite{zan1}. This model is exactly solvable, so as a curiosity we check what our RG method predicts as the information metric. A bosonization
procedure near $\gamma = 0$, $h=0$ here yields the RG equations ${\dot \gamma} = \gamma$, ${\dot h} = h$. Along with Eq.(\ref{metriclin}) and
the fact that for linearized RG flows, the information metric can be taken to be diagonal \cite{diag}, 
we recover $g_{\gamma\gamma} \sim \gamma^{-1}$, $g_{hh} \sim \gamma^{-1}$, in agreement to the metric derived in Eq.(7) of \cite{zan1}, near
$\gamma = 0$. 

\section{Nonlinear Examples and Logarithmic Corrections}
In this section, we will focus on cases where the RG flow equations include nonlinear higher order terms. As in the previous section, an 
appropriate combination of variables will be seen to render the solutions tractable. 

\subsection{Case IV}

Our first example is given by the flow equations
\begin{equation}
{\dot x} = a_1x + a_2x^2,~~~{\dot y} = b_1y 
\label{nonlin0}
\end{equation}
These flows arise, for example, in the critical dynamics of a time independent random field interaction introduced in a Ising spin or quantum rotor 
model \cite{Dutta1} in the $\epsilon$ expansion, above the lower critical dimension $d_c=2$. The Hamiltonian for the model is 
\begin{equation}
H = -J\sum _{\langle ij \rangle} S_i^z S_i^z - \Gamma \sum S_i^x - \sum_i h_i S_i^z
\label{DuttaStinch}
\end{equation}
where $\Gamma$ is the strength of a transverse field, and $h_i$ are site dependent magnetic fields, with the random fields correlated
in one direction. RG flows in this model were studied in \cite{Dutta1} and the equations are of the form of Eq.(\ref{nonlin0}), with the
identification $x = h/J$ and $y = T/J_0$ where $T$ is the temperature, $h$ is a measure of the randomness of the random field $h_i$ defined via its
distribution (Eq.(2) of \cite{Dutta1}), $J_0$ is the interaction in the direction in which the fields are correlated. Also, for this model,
$a_1 = -\epsilon/2$ and $b_1 = -(1+\epsilon)$, with $d = 2+\epsilon$. In general, obtaining the information metric for such a system is not 
possible exactly, but can be done using the symmetry of the RG flow equations, as we
show below. 

To keep the discussion general, we proceed with arbitrary $a_1$, $a_2$ and $b_1$. First, note that a transformation of variables 
$z = x/(a_1 + a_2x)$ renders the equations linear in the variables $z$ and $y$, i.e ${\dot z} = a_1z$ and ${\dot y} = b_1y$. Now we can
construct the information metric using the methods described in the previous subsection. Since both $x$ and $y$ are irrelevant up to linear order, i.e are
stable directions, following our previous discussion the scalar curvature does not blow up here. This is the trivial fixed point. 
In order to obtain the information metric near the non-trivial fixed point, it is more convenient to work in terms of the original variables, and we
illustrate the results below. The differential equations for the metric components are obtained in the variables $x$ and $y$ as 
\begin{eqnarray}
&~& 2(a_1 + 2a_2x)g_{xx} + b_1yg_{xx,y} + x(a_1 + a_2x)g_{xx,x} - Dg_{xx} = 0,\nonumber\\
&~& (a_1 + b_1 + 2a_2x)g_{xy} + b_1yg_{xy,y} +x(a_1 + a_2x)g_{xy,x} - Dg_{xy}=0,\nonumber\\
&~& 2b_1g_{yy} + b_1yg_{yy,y} + x(a_1 + a_2x)g_{yy,x} - Dg_{yy} = 0.
\end{eqnarray}
These have the solutions given by 
\begin{eqnarray}
g_{xx} &=& x^{\frac{D}{a_1}-2} \left(a_2 x+a_1\right){}^{-\frac{D}{a_1}-2},~~g_{xy}=x^{-\frac{a_1+b_1-D}{a_1}} \left(a_2 x+a_1\right){}^{-\frac{a_1-b_1+D}{a_1}},\nonumber\\
g_{yy} &=& \left(\frac{x}{a_2 x+a_1}\right){}^{\frac{D-2 b_1}{a_1}},
\end{eqnarray}
The geodesic equations can be obtained in the same way as outlined in the previous section, and denoting the tangent vector as 
$(x'(\lambda), y'(\lambda))$, we obtain as a normalized solution 
\begin{equation}
x'(\lambda) = \sqrt{2}x(a_1+a_2x)\left(\frac{x}{a_1 + a_2x}\right)^{-\frac{D}{2a_1}},~ 
y'(\lambda) = -\frac{1}{\sqrt{2}}\left(\frac{x}{a_1 + a_2x}\right)^{-\frac{D}{2a_1}+\frac{b_1}{a_1}}
\end{equation}
The above equation then yields 
\begin{equation}
\lambda = \frac{\sqrt{2}}{D}\left(\frac{x}{a_1 + a_2x}\right)^{\frac{D}{2a_1}}
\end{equation}
Noting that the scalar curvature and the expansion parameter are given by  
\begin{equation}
R = 2 b_1 \left(D-2 b_1\right) \left(\frac{x}{a_2x+a_1}\right){}^{-\frac{D}{a_1}},~~~
\Theta = \frac{D - 2b_1}{\sqrt{2}}\left(\frac{x}{a_2 x+a_1}\right){}^{-\frac{D}{2 a_1}}
\end{equation}
we obtain $R \sim \lambda^{-2}$, $\Theta \sim \lambda^{-1}$, as expected. Note that these are true in $D = 2+\epsilon$ dimensions. 
Since $a_1 = -\epsilon/2$ is negative, the curvature and the expansion parameters diverge at the non-trivial fixed point $x^* = -a_1/a_2$ 
and $t^* = 0$. There is no divergence of these quantities at the trivial fixed point $x^* = t^* = 0$. 

\subsection{Case V}

Next, we come to the case where the RG equations are taken to be 
\begin{equation}
{\dot x} = a_1x + a_2x^2,~~~{\dot y} = b_1y + b_2xy
\label{nonlin1}
\end{equation}
These is the standard textbook form for RG equations of Ising like models near four dimensions, i.e in $d = 4 - \epsilon$, if we identify $a_1 = \epsilon$,
$a_2 = -72$, $b_1 = 2$, $b_2 = -24$. For $O(N)$ vector models, the coefficients $a_2$ and $b_2$ are given by $-8(n+8)$ and 
$-8(n+2)$ respectively \cite{Cardy}. 

First note that upon making the coordinate transformations $z_1 = x/(a_1 + a_2x)$, $z_2 = y(a_1 + a_2x)^{-\frac{b_2}{a_2}}$, these equations 
reduce to ${\dot z_1} = a_1z_1$, ${\dot z_2} = b_1z_1$. This clearly defines the Gaussian fixed point, with the critical exponent along the $y$
direction being $b_1=2$. Next, if we linearize the RG equations near the Wilson-Fisher fixed point $x^* = -a_1/a_2 = \epsilon/72$,
we get back the thermal exponent $2 - \epsilon/3$, as follows from our discussion of the previous section. 

It is interesting to see what the information metric reveals when applied to non-linear set of equations, Eq.(\ref{nonlin1}). Here, 
it is convenient to first define a new variable $z = y(a_2 + a_1/x)^{b_1/a_1}$, so that we have an equivalent set of equations
\begin{equation}
{\dot x} = a_1x + a_2x^2,~~~{\dot z} = b_2xz
\label{WF1}
\end{equation}
Note that since $b_1$ and $\epsilon$ are positive, the coordinate transformation mentioned above becomes ill defined near $x\to 0$, i.e 
the Gaussian fixed point. However, this is not the case near the Wilson-Fisher fixed point. We will need this fact later. 
We take the components of the homothetic vector field, generating the scale transformation near criticality as $({\dot x}, {\dot z})$. 
Then, the equations determining the metric on the parameter manifold 
(with parameters $(x,z)$) are given by 
\begin{eqnarray}
&~&x\left(2b_2g_{zz} + b2zg_{zz,z} + (a_1 + a_2x)g_{zz,x}\right) - Dg_{zz}=0~,\nonumber\\
&~&\left(a_1+x(2a_2 + b_2\right)g_{xz} + b_2zg_{zz} + x(b_2yg_{xz,z} + (a_1+a_2x)g_{xz,x}) - Dg_{xz}=0~,\nonumber\\
&~& 2(a_1+2a_2x)g_{xx} + 2b_2zg_{xz} + x\left(b_2zg_{xx,z} + (a_1 + a_2x)g_{xx,x}\right) - Dg_{xx}=0.
\end{eqnarray}
The metric can be determined from the above equations by first solving for $g_{zz}$ and hence $g_{xz}$ and $g_{xx}$. The expressions are lengthy, but simplify in the
limit $z=0$ (equivalently $y=0$) and read
\begin{eqnarray}
g_{xx} &=& x^{\frac{D}{a_1}-2} \left(a_2 x+a_1\right){}^{-\frac{D}{a_1}-2}~,~~~
g_{xz} = x^{\frac{D}{a_1}-1} \left(a_2x+a_1\right){}^{-\frac{b_2}{a_2}-\frac{a_1+D}{a_1}}~,\nonumber\\
g_{zz} &=& x^{\frac{D}{a_1}} \left(a_2 x+a_1\right){}^{-\frac{2b_2}{a_2}-\frac{D}{a_1}}.
\label{near4Dmetric}
\end{eqnarray}

Having obtained the metric, we now focus on geodesics on the PM. As usual, we start with the tangent vector $(x'(\lambda), z'(\lambda))$. The fact that
the metric is independent of $z$ leads to 
\begin{equation}
x'(\lambda) = \sqrt{2}x^{1-\frac{D}{2a_1}}\left(a_1+a_2x\right)^{1+ \frac{D}{2a_1}},~
z'(\lambda) = -\frac{1}{\sqrt{2}}x^{-\frac{D}{2a_1}}\left(a_1 + a_2x\right)^{\frac{D}{2a_1} + \frac{b_2}{a_2}}
\end{equation}
The first of the above equations can then be solved to give
\begin{equation}
\lambda = \frac{\sqrt{2}}{D}\left(\frac{x}{a_1+a_2x}\right)^{\frac{D}{2a_1}}
\label{lambda1}
\end{equation}
which was also the expression obtained for the quantum rotor model studied in subsection 3.2. The scalar curvature and the expansion parameter are given by 
\begin{eqnarray}
R &=&  -2 b_2 x^{1-\frac{D}{a_1}} \left(a_2 x+a_1\right){}^{\frac{D}{a_1}} \left(-2a_2 x-2 a_1+2 b_2 x-D\right)\nonumber\\
\Theta &=& \frac{1}{\sqrt{2}}\left(D - 2b_2x\right)\left(\frac{x}{a_1+a_2x}\right)^{-\frac{D}{2a_1}}
\label{WF}
\end{eqnarray}
Now note that in this case, $a_1 = \epsilon $ is positive. Hence, the scalar curvature and the expansion parameter seems to diverge 
at the Gaussian fixed point $x\to 0$. However, as mentioned before, the coordinate transformation used to derive Eq.(\ref{WF1}) is not trustable here. 
Let us thus focus on the Wilson-Fisher fixed point, where from Eq.(\ref{lambda1}), we also see that $\lambda \to \infty$.  This may look at variance
with the linearized result (where the geodesic distance goes to zero near a non-trivial fixed point), but that is not so. To see this note that in 
our non-linear analysis, we have not linearized about the Wilson-Fisher point. The geodesic distance here is still measured from $x=y=0$, so our 
analysis simply reflects the fact that the Wilson-Fisher fixed point is an infra-red fixed point. 
By combining Eqs.(\ref{lambda1}) and (\ref{WF}), we recover the relations $R \sim \lambda^{-2}$ and $\Theta \sim \lambda^{-1}$ at this fixed point. 

\subsection{Case VI }

Another interesting situation occurs when one of the variables is marginally irrelevant. We will focus on a set of RG flow equations
of the form 
\begin{equation}
{\dot x} = a_1 x^2,~~~{\dot y} = b_1 y + b_2 x y
\label{4dRG}
\end{equation}
These are the RG flow equations for the Ising model in four dimensions (with $a_1=-72$, $b_1=2$ and $b_2=-24$), 
as follows from our analysis of the previous section, by setting $\epsilon = 0$. Clearly, the results of that analysis cannot be used 
here simply by setting the coefficient of the linear term to zero, since the metric components of 
Eq.(\ref{near4Dmetric}) are ill defined in this limit. To perform this analysis, we start by defining a new variable $z =  ye^{b_1/(a_1x)}$. 
Since $a_1$ is taken to be negative, and hence 
this transformation is well defined in the limit $x \to 0$. It follows that the RG equations can be written more conveniently as 
\begin{equation}
{\dot x} = a_1 x^2,~~~{\dot z} = b_2 x z
\label{4dRG}
\end{equation}
In terms of the variables $x$ and $z$, the equations for the metric components in this case are seen to be 
\begin{eqnarray}
&~&x\left(2b_2g_{zz} + b_2zg_{zz,z} + a_1xg_{zz,x} \right) -Dg_{zz} = 0~,\nonumber\\
&~&(2a_1+b_2)xg_{xz} + b_2zg_{zz} + x\left(b_2zg_{xz,z} + a_1xg_{xz,x}\right) -D g_{xz} = 0~,\nonumber\\
&~&4a_1xg_{xx} + 2b_2zg_{xz} + x\left(b_2zg_{xx,z} + a_1xg_{xx,x}\right) + Dg_{xx} =0~.
\end{eqnarray}
As before, the first of these equations can be solved to obtain $g_{zz}$, which can in turn be used to find $g_{xz}$ and hence $g_{xx}$. The solutions for the metric components read
\begin{eqnarray}
g_{xx}&=&\frac{1}{a_1^2}x^{-\frac{2 b_2}{a_1}-4} e^{-\frac{D}{a_1 x}} \left(a_1 x^{\frac{b_2}{a_1}}-b_2 xz\right)^2,~
g_{xz} = \frac{1}{a_1}x^{-\frac{2 (a_1+b_2)}{a_1}} e^{-\frac{D}{a_1 x}} \left(a x^{\frac{b_2}{a_1}}-b_2 xz\right),\nonumber\\
g_{zz} &=& k_1 x^{-\frac{2 b_2}{a_1}} e^{-\frac{D}{a_1 x}}
\label{metlog}
\end{eqnarray}
Here $k_1$ is a constant that we will set to $2$ following our previous discussions. 
We first record the expression for the scalar curvature calculated from the metric of Eq.(\ref{metlog}) in the limit $z=0$,
\begin{equation}
R =\frac{2 b_2 x e^{\frac{D}{a_1 x}} (2 a_1 x-2 b_2 x+D)}{a_1^2}
\label{Rlog}
\end{equation}
Clearly, with negative $a_1$, $R$ rapidly goes to zero for small values of $x$.  
Now let us consider geodesics on the manifold defined by the metric of Eq.(\ref{metlog}). For the four dimensional Ising model, $a_1 = -72$ and 
$b_2 = -24$, and for small $x$ and $z$, we can ignore the $xz$ pieces in Eq.(\ref{metlog}), which renders the said metric independent of $z$. 
As before, we consider a tangent vector to a geodesic, which we denote by $(x'(\lambda), z'(\lambda))$. Here, $\lambda$ is an affine parameter 
along the geodesic. Normalization of the tangent vector, along with the fact that the metric does not depend on $z$ implies that
\begin{equation}
x'(\lambda )= \sqrt{2} x(\lambda )^2 e^{\frac{D}{2 a_1 x(\lambda )}},~~~
z'(\lambda ) =-\frac{1}{\sqrt{2}} x(\lambda )^{\frac{b_2}{a_1}} e^{\frac{D}{2 a_1 x(\lambda )}}
\end{equation}
Solving the first of these equations, we obtain
\begin{equation}
x(\lambda )= -\frac{D}{2 a_1 \log \left(\frac{-D \lambda }{\sqrt{2}a_1}\right)}.
\end{equation}
If $a_1$ is negative, as happens for the four dimensional Landau-Ginzburg model, we see that at $x=0$, $\lambda$ has to go to infinity,
although logarithmically \cite{log}. After some algebra we obtain here, 
\begin{equation}
R=-\frac{2 b_2 \left(a_1 \log \left(-\frac{D \lambda }{\sqrt{2}a_1}\right)-a_1+b_2\right)}{a_1^2 \lambda ^2 \log ^2\left(-\frac{D \lambda}{\sqrt{2} a_1}\right)},~~
\Theta = \frac{1}{\lambda}\left(1+\frac{b_2}{a_1 \log \left(-\frac{D \lambda }{\sqrt{2}a_1}\right)}\right)
\label{logcorr}
\end{equation}
In the limit $\lambda \to \infty$, we finally obtain
\begin{equation}
R \sim \frac{1}{\lambda^2{\rm log}\lambda},~~\Theta \sim \frac{1}{\lambda}
\end{equation}
The first of these equations indicate that the Ricci scalar picks up logarithmic corrections to geometric scaling relations in four dimensions. 

\section{Conclusions}

In this paper, we have provided evidence that scale invariance in the vicinity of a critical point can provide valuable information on the metric of
the parameter manifold in classical and quantum phase transitions, in a unified fashion. In particular, this method can be applied to systems that are not exactly solvable 
to read off the scaling behavior of the metric (and hence related quantities like the fidelity susceptibility) near criticality. Our method complements
the work of \cite{VenZan}) to determine scaling patterns for information geometric quantities. While the work of \cite{Diosi} utilises the RG equations up to 
first order to read off the scaling of the metric in classical phase transitions,
our method explicitly solves for the metric components in a variety of non-linear examples. 

While most of the paper deals with classical phase transitions, we have given one non-trivial example of a quantum phase transition that can be
studied in this framework. This extends the study of information geometry to 
novel settings, which, to the best of our knowledge, have not appeared in the literature.
We have seen here that the relations $R \sim \lambda^{-2}$ and $\Theta \sim \lambda^{-1}$ are universal, except for the four dimensional Ising
model, where the former relation picked up logarithmic corrections. This strengthens the claim made in \cite{tapo3} about universal geometric
critical exponents. 

We should mention here that in two dimensions, the scalar curvature $R$ and the expansion parameter $\Theta$ satisfy the Raychaudhuri equation 
$\Theta^2 + \Theta' + R/2=0$, where a prime denotes a derivative with respect to the affine parameter $\lambda$ \cite{reason}. In principle, given $R$ ($\Theta$), 
this equation can be used to determine $\Theta$ ($R$). However, it is not always possible to solve this equation analytically. In all the examples
considered in this paper, as a cross check on our results, we have verified that the Raychaudhuri equations are indeed satisfied. 

In this paper, we have considered a class of examples where the information metric was obtained from the set of RG flow equations. Clearly, one
might argue that this may not be the case for more generic examples. Consider for example an RG equation of the form 
${\dot x} = a_1x + a_2y$, ${\dot y} = b_1y + b_2x$. In this case, the homothetic vectors do not have an analytic solution as can be checked. 
This is a caveat in our analysis. A further criticism might be that higher order terms than those considered here, are difficult to take care of. 
We note however that in general, such terms might be solved iteratively, i.e one can use perturbation theory to solve the 
differential equations for the components of the metric tensor. This is substantially more complicated that the analysis presented here, 
and a full study of the same is left for the future. 

It will be interesting to extend the present analysis to cases where the parameter manifold has dimensionality higher than two. One example was
commented upon in this work, but a broader analysis might reveal interesting facts about the geometry of the renormalization group, as higher
dimensional PMs offer more structure, and in particular, more scalar invariants. What these scalars mean in the context of RG will be an interesting
issue for future investigation. It might also be interesting to consider the role of time in information theory \cite{Anandan}, in the context of
the models considered here. This issue is currently under investigation. Finally, it might be useful to investigate information geometry in the 
context of Kosterlitz-Thouless type phase transitions in the two dimensional $XY$ model. Preliminary analysis indicates that here the scalar 
curvature of the information metric diverges exponentially, in lines with the behavior of the correlation length. However, this case requires further
understanding. 

\begin{center}
{\bf Acknowledgements}
\end{center}
It is a pleasure to thank Amit Dutta for numerous fruitful discussions. The work of SM is supported by grant no. 09/092(0792)-2011-EMR-1 from CSIR, India.


\begin{thebibliography}{99}
\bibitem{curvature}
For a two dimensional manifold with coordinates $(x^1,x^2)$,
$R = \frac{2}{\sqrt{g}}\left[\frac{\partial}{\partial x^2}\left(\frac{\sqrt{g}}{g_{11}}\Gamma^2_{11}\right) -\frac{\partial}{\partial x^1}\left(\frac{\sqrt{g}}{g_{11}}\Gamma^2_{12}\right)\right]$,
where $g$ is the determinant of the metric, and the Christoffel symbols are defined by 
$\Gamma^{\mu}_{\nu\rho} = \frac{1}{2}g^{\mu\zeta}\left(\frac{\partial g_{\zeta\nu}}{\partial x^{\rho}} + \frac{\partial g_{\zeta\rho}}{\partial x^{\nu}}
- \frac{\partial g_{\nu\rho}}{\partial x^{\zeta}}\right)$. For higher dimensional manifolds, the formula is standard but more complicated. See, e.g \cite{poisson}.
\bibitem{geofocus}
The converse is not true : geodesic focusing does not always indicate a singularity. This is seen for example, by considering a two-sphere, 
where great circles focus at the poles which are regular points on the sphere.
\bibitem{diag}
In this particular example, we could also have begun with a diagonal metric with $g_{xy}$ set to zero. This choice also gives 
a consistent set of differential equations for the metric and leads to the same conclusions as with a generic off-diagonal form of the metric. However,
we proceed with a generic off diagonal metric, to keep the discussion general, and to connect to a known example later in this subsection.
\bibitem{Hom1}
In the examples that we study in sequel, the components of the homothetic vector field are obtained from the RG equations in the same way. These are then
put in Eq.(\ref{homothety}) and equations for the metric components are obtained.
\bibitem{aneg}
There is a small subtlety here. In this example, if $a$ is negative (and in some other situations that we come across later), setting $k_3$ to zero 
indicates that  $\lambda \to \infty$ as $x \to 0$. In this example, we will interpret this as a mathematical artefact indicative of a stable fixed point.
\bibitem{finetune}
In this case, it is possible to adjust the constant $k_3$ appearing in Eq.(\ref{lam1}) to very large values to keep the value of $\lambda$ close to zero. 
But this requires fine tuning of the constant $k_3$. In what follows, we will assume that $\lambda$ becomes large along an irrelevant (stable) direction.
\bibitem{reason} In two dimensions, the components of the Ricci tensor are related to the metric, 
$R_{\mu\nu} = \frac{1}{2}g_{\mu\nu}R$, with $R$ being the Ricci scalar. This is the required ingredient in this derivation.
\bibitem{log}
If on the other hand we wanted $\lambda \to 0$, then $x$ becomes negative, which renders the metric components of Eq.(\ref{metlog}) 
to be imaginary, which we should avoid, from physical considerations.
\end{thebibliography}
\end{document}